\documentclass[twocolumn,showpacs,aps,pra]{revtex4}
\usepackage{graphics,epsfig,graphicx}
\usepackage{color}
\usepackage{makeidx}
\usepackage{latexsym}

\def\ket#1{\mathinner{|{#1}\rangle}}

\begin{document}

%======================================%
%<<<<<<<<<<<< TITLE PAGE >>>>>>>>>>>>>>%
%======================================%

\title{Quantum metrology to probe atomic parity nonconservation}
\author{P.~Mandal\footnote{drupm@iacs.res.in} and M.~Mukherjee}
\affiliation{Raman Center for Atomic Molecular and Optical Sciences,
Indian Association for the Cultivation of Science, Kolkata -700 032}

\date {Received {\today}}

\begin{abstract}

An entangled state prepared in a decoherence free sub-space together
with a Ramsey type measurement can probe parity violation in heavy
alkali ions like Ba$^+$ or Ra$^+$. Here we propose an experiment
with Ba$^+$ as an example to measure the small parity violating
effect in this system. It has been shown that a measurement on a
maximally correlated system will reduce the uncertainty as compared
to that on a single ion measurement. In addition it also provides a
feasible solution to measure the nuclear spin dependent part of the
total parity violating light shift in an ionic system which has so
far not been addressed.

\end{abstract}

\pacs{03.65.Ta,  42.50.Lc  32.80.-t, 42.50.Ct, 42.50.Vk, 32.80.Lg }

\maketitle

%======================================%
%<<<<<<<<<<<<<<< TEXT >>>>>>>>>>>>>>>>>%
%======================================%

Measurement of atomic parity nonconservation (PNC) in the $6S-7S$
transition of atomic Cs has been performed with an uncertainty
reaching $0.35\%$~\cite{Woo97,Ben99,Bou97}. An equally demanding
theoretical effort in this atom~\cite{Gin04} leads to the evaluation
of the weak nuclear charge $Q_W$ which is a unique low energy test
of the standard electroweak theory. Further improvement in the
precision will lead to reducing the limits on the mass of an
eventually additional light or heavy boson~\cite{Bou05}. Apart from
the necessity of improving the PNC measurement in Cs, it would be
worthwhile to consider other possible experimental techniques for
the measurement of PNC in other systems. Recently the largest PNC
effect has been measured in the $6s^{2}$~$^{1}S_{0}-5d6s^{3}D_{1}$
transition in atomic yetterbium~\cite{tsi09} employing the same
technique as used in the Cs experiment. The enhancement in this case
is caused by degeneracy of atomic levels~\cite{dem95}. Though the
measured PNC dipole transition amplitude~($E1_{PNC}$) is $100$ times
larger compared to that in atomic cesium, the experimental precision
is not good enough to verify the Standard Model or to predict any
Physics beyond it.

A newly proposed method adopted  for  the  Cs measurement involved
left-right asymmetry of the forbidden transition rate in $6S-7S$
transition~\cite{Gue03,Gue05}. This method is presently been pursued
for Fr, the heaviest alkali~\cite{Gom06,Aub03}. Unfortunately, the
requirement of a large number of atoms  to observe the asymmetry
limits this experiment. Recently, it  has been proposed to  observe
a linear  Stark shift in an interferometric measurement with small
number of atoms of Fr~\cite{Bou08}. The measurement of light shift
arising due to the interference between $E1_{PNC}$ and electric
quadrupole transition amplitude~$E2$ in a heavy ion like Ba$^+$,
Ra$^+$ proposed by Fortson~\cite{Nov93} seems to be the most
promising technique. It can, in principle achieve a precision of
$0.1$\%. Presently it is being pursued at different experimental
laboratories~\cite{Ver10,Man10}. Initial radio frequency~(RF)
spectroscopy on Ba$^+$ has also been performed to observe the light
shifts of different Zeeman sublevels. The major limitations appear
in these measurements are from the magnetic field noise as well as
laser frequency noise~\cite{She05}. In order to finally observe the
PNC induced light shift it is necessary to achieve an uncertainty
well below one Hertz in the ground state Larmor frequency since even
in the presence of a strong electric field, the shift is only of the
order of $0.2~$Hz. Though, maximally entangled states for quantum
metrology is a rather recent field, it has already been implemented
in a relatively few cases~\cite{Bol96,Gio04,Gio06}. They have been
used to improve the signal-to-noise ratio~\cite{Lei04}, to
efficiently detect quantum state~\cite{Sch05}, to measure scattering
length~\cite{Wid04} and to do spectroscopy in decoherence free
sub-space~(DFS)~\cite{Roo06}. An entangled state prepared in a
DFS~\cite{Lid98,Roo04} makes any measurement immune to environmental
changes. Thereby, this can be effectively used to overcome the
magnetic field noise limitation of the single ion experiment to
observe the PNC light shift. In what follows, we outline this
promising technique of such a measurement with high precision.

Parity nonconservation in an atomic system leads to a small mixing
between states of opposite parities resulting a nonzero probability
in the electric dipole transition which is strictly forbidden by
parity conservation rule. The effect, though scales as $Z^{3}$ for
heavier atoms~\cite{Bou97}, is on the order of $10^{-11}ea_{0}$. It
is thus an experimental challenge to measure such a small quantity
directly. Instead, people look for an interference-like phenomena
between $E1_{PNC}$ and a much stronger higher order electromagnetic
transition between the same states. For Ba$^+$ or Ra$^+$ such an
interference between $E1_{PNC}$ and $E2$ in
nS$_{1/2}-$(n-1)D$_{3/2}$ transition is proposed to measure the
vector light shift~\cite{Nov93}. In presence of the electric field
of a laser
\begin{equation}
\label{eqn1}
\textbf{E}(\textbf{r},t)=\frac{1}{2}\textbf{E}_{0}[e^{i(\textbf{k}.\textbf{r}
- \omega t)} + c.c.],
\end{equation}
the $E1_{PNC}$ and $E2$ couplings between S$_{1/2}$$-$D$_{3/2}$ are
described in terms of respective Rabi frequencies as
\begin{eqnarray}
\label{eqn2}
\Omega^{PNC}_{m'm}=\frac{1}{2\hbar}\sum_{i}\varepsilon^{PNC}_{m'm}E_{i}(0) \\
\Omega^{Q}_{m'm}=\frac{1}{2\hbar}\sum_{i,j}\varepsilon^{Q}_{m'm}[\frac{\partial
E_{i}(r)}{\partial x_{j}}]_{0},
\end{eqnarray}
$r=0$ being the position of the ion in the trap. Here
$\varepsilon^{PNC}_{m'm}$ and $\varepsilon^{Q}_{m'm}$ describe
$E1_{PNC}$ and $E2$ matrix elements between $m$ sublevel of
S$_{1/2}$ and $m'$ sublevel of D$_{3/2}$. The resultant Rabi
frequency of $m$ sublevel of S$_{1/2}$ is~\cite{Nov93}
\begin{equation}
\label{eqn3}
\Omega_{m} \approx \Omega^{Q}_{m} +
 Re\sum_{m'}(\Omega^{PNC\ast}_{m'm}\Omega^{Q}_{m'm})/\Omega^{Q}_{m},
\end{equation}
where $(\Omega_{m})^{2} = \sum_{m'}|\Omega_{m'm}|^{2} =
\sum_{m'}|\Omega^{Q}_{m'm} + \Omega^{PNC}_{m'm}|^{2}$ and
$(\Omega^{Q}_{m})^{2} = \sum_{m'}|\Omega^{Q}_{m'm}|^{2}$.
Considering the Zeeman splitting of the magnetic sublevels to be
comparable to the line width of S$_{1/2}$$-$D$_{3/2}$ laser, the
light shift of $m$ sublevel of the ground state is given by
\begin{equation}
\label{eqn4}
\Delta\omega_{m} = \delta/2 - \Omega_{m}
\end{equation}
where $\delta = \omega_{0} - \omega$ is the detuning of the laser
frequency from the atomic transition frequency. It is convenient to
drive both the quadrupole and PNC allowed S$_{1/2}$$-$D$_{3/2}$
dipole transition independently so that much larger contribution in
$\Delta\omega_{m}$ due to the pure $E2$ coupling remains same while
that due to the interference term changes sign for the magnetic
sublevels of ground state. Fortson showed~\cite{Nov93} that it is
achieved when a single ion is placed ($x=z=0$) simultaneously at the antinode
and the node of two standing wave lasers represented respectively as
\begin{equation}
\label{eqn5}
E'=\hat{x}E'_{0}\cos{kz}, E''=i\hat{z}E''_{0}\sin{kx}.
\end{equation}
These lasers produce $\Delta m = \pm 1$ dipole and
quadrupole transitions respectively. In presence of these two lasers
the Larmor frequency between the ground magnetic sublevels is given
by
\begin{equation}
\label{eqn6} \omega'_{L} = \omega_{L} \sim
2Re\sum_{m'}(\Omega^{PNC\ast}_{m'm}\Omega^{Q}_{m'm})/\Omega^{Q}_{m},
\end{equation}
where $\omega_{L}$ is the Larmor frequency between the same
sublevels in absence of the lasers. Thus the PNC shift can be extracted
from the measurement of the ground state Larmor frequency in absence
and in presence of the laser fields. Fortson calculated the shift to
be $0.2$ Hz for Ba$^{+}$ in presence of strong laser field $E'_{0} =
2 \times 10^{6}$~V/m~\cite{Nov93}. However, it is still a challenge
to measure such a small change by applying usual RF spectroscopic
technique. It demands a magnetic field of stability one part in
$10^{8}$ for few hundreds of a kHz magnetic splitting in order to
achieve an accuracy $1~\%$.

Employing the generalized Ramsey interferometric technique to
maximally correlated atomic state it is possible to determine the
PNC light shift with the required precision. Under free precision a
maximally entangled atomic state, like one of the Bell's states,
$\psi(0)=\frac{1}{\sqrt{2}}(\ket{u_1}\ket{u_2}+\ket{v_1}\ket{v_2})$
evolves into $\psi(\tau)=\frac{1}{\sqrt{2}}(\ket{u_1}\ket{u_2}+
\exp^{i\Delta \lambda \tau}\ket{v_1}\ket{v_2})$ after a time $\tau$.
The phase evolution rate $\Delta \lambda =
[(E_{u_1}+E_{u_2})-(E_{v_1}+E_{v_2})]/\hbar$ corresponds to the
energy difference between the atomic states $u_k$ and $v_k$. The
real part of the phase factor $\exp^{i\Delta \lambda t}$ can be
measured by projecting the ions on the states
$\ket{\pm}=\frac{1}{\sqrt2}(\ket u_k \pm \ket v_k)$ and measuring
the relative phase. For states in the DFS the free precision time
$\tau$ can be made very long and hence the phase can be measured
accurately~\cite{Roo04,Hae05}. By a careful choice of the state it
is possible to measure the PNC shift in DFS thereby avoiding the
possible systematic effects in coupling to the environment.

\begin{figure}
  \includegraphics[width=7cm]{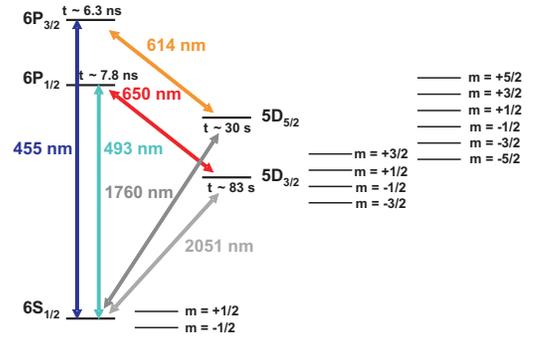}\\
  \caption{(Color online) Relevant atomic levels of Ba$^+$. The Zeeman sublevels are also shown for clarity.}\label{Fig1}
\end{figure}

Instead of a single ion, in the following we consider a string of
two Ba$^+$ ions (even isotope, $I=0$) confined in a linear Paul
trap. The relevant electronic levels are shown in fig.~\ref{Fig1}.
The ions cooled into their ground motional state of the first two
normal modes of motion~\cite{Roo99} using laser Doppler cooling
applying $493$~nm and $650$~nm lasers followed by sideband cooling
with $1.76$$~\mu$m laser~[fig.~\ref{Fig1}]. Both the ions are
prepared in a Zeeman sublevel of the ground electronic state (say
$6S_{1/2}, m=1/2$ for example). The ions are then individually
treated with a sequence of laser pulses. A $\pi/2$ pulse at the blue
sideband on the first ion prepares it in a superposition of
electronic ground and metastable (say $D_{5/2}, m=1/2$ say for
example) state and motional ground and first excited state. A $\pi$
pulse at the carrier on the second ion brings it to electronic
excited state ($D_{5/2}, m=1/2$) keeping the motional state
unchanged. Another blue sideband $\pi$ pulse on the second ion
transfers the excited state population back to ground electronic and
motional states. One more $\pi$ pulse at the carrier on each ion
coherently transfers the quadrupole excited state population into
other Zeeman sublevel in the ground state thus preparing the state
\begin{equation}
\label{eqn7} |\Psi\rangle =
\frac{1}{\sqrt{2}}(|1\rangle_{1}|0\rangle_{2} +
|0\rangle_{1}|1\rangle_{2})|0\rangle,
\end{equation}
where $|1\rangle_{i}$ and $|0\rangle_{i}$ stand for $m=1/2$ and
$-1/2$ of 6S$_{1/2}$ state of $i^{th}$ ion and $|0\rangle$ describes
the ground motional state of center-of-mass~(COM) mode. The presence
of the two ion state makes it decoherence free as compared to the
superposition state of a single ion. The Zeeman shifts of the two
parts of the entangled state cancel out in absence of magnetic field
gradient along the trap axis. This state is immune to any
decoherence effects arising from the magnetic field fluctuation
common to both ions, spontaneous decay \emph{etc} and therefore the
state, in principal possesses an infinitely long coherence.

\begin{figure}
  \vspace{-1cm}
  \includegraphics[width=6cm]{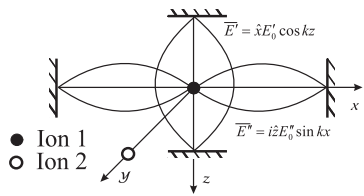}\\
  \caption{A schematic of the experiment with two ions placed in a linear ion trap and interrogated by two standing wave
  lasers. The amplitude $E'_0$ should be orders of magnitude larger as compared to $E''_0$ for improved systematic.}\label{Fig2}
\end{figure}

\begin{figure}
  \includegraphics[width=5cm]{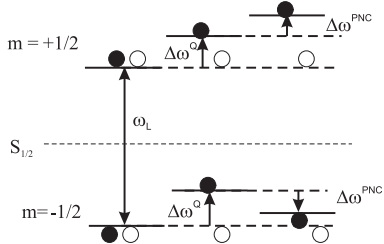}\\
  \caption{A cartoon of energy shifts of ground state magnetic sublevels of two ions
  in presence of the magnetic field and lasers $E'$, $E''$. $\Delta\omega^{Q}$ and
  $\Delta\omega^{PNC}$ denotes quadrupole and PNC light shifts respectively.}\label{Fig3}
\end{figure}

After preparing such an entangled state in DFS, two laser fields
$E'$ and $E''$ in a standing wave configuration are applied for a
time interval $\tau$ on one ion (Ion 1, say) as shown in
fig.~\ref{Fig2}. The magnetic splitting, quadrupole light shift and
PNC light shift of the ground state magnetic sublevels for the two
ions are shown schematically in fig.~\ref{Fig3}. It depicts that the
ground state Larmor frequency of one ion shifts only due to PNC
interaction while that remains unchanged for the other ion. Thus a
small perturbation is introduced within the entangled state
$|\Psi\rangle$~(Eq.~\ref{eqn7}) and it evolves as
\begin{equation}
\label{eqn8} |\Psi(\tau)\rangle =
\frac{1}{\sqrt{2}}(|1\rangle_{1}|0\rangle_{2} +
\exp(i\Delta\lambda\tau) |0\rangle_{1}|1\rangle_{2})|0\rangle,
\end{equation}
where the phase evolution rate $\Delta\lambda$ corresponds to the
energy difference between the two parts of the entangled state
\emph{i.e.} the PNC light shift given by Eq.~\ref{eqn6}.

The state $|\Psi(\tau)\rangle$ should be projected on
$\ket{\pm}=\frac{1}{\sqrt2}(\ket 0_i \pm \ket 1_i)$ in order to
observe the time evolution of the expectation value
$\langle\sigma^{(1)}_{x}\otimes\sigma^{(2)}_{x}\rangle$, where
$\sigma^{(i)}_{x}$ denotes the Pauli spin matrix for $i^{th}$ ion.
It oscillates with a frequency $2\pi/\Delta\lambda$~\cite{Roo04}.
Thus the PNC light shift can directly be extracted from the
measurement of the oscillation frequency.

The uncertainty in the PNC light shift measurement will be
determined by the decoherence time of the maximally entangled state
which is practically infinite in absence of external perturbation but limited by
the natural life time ($\tau$) of $5D_{3/2}$ state in our case. The
uncertainty in the frequency measurement on $N$ maximally correlated
atomic systems is inversely proportional to $NT$ instead of
$\sqrt{N}T$ for uncorrelated systems~\cite{Bol96}. Here $T$ is the
time of a single measurement which can be made as large as $\tau$
is. Therefore the statistical signal-to-noise ratio for $n$ no. of
measurements can be approximated as
\begin{equation}
\frac{\varepsilon^{\text{PNC}}}{\delta
\varepsilon^{\text{PNC}}}\approx
\frac{\varepsilon^{\text{PNC}}E'_0}{\hbar}f\sqrt{n}N\tau,
\end{equation}
where $f$ signifies an experimental efficiency factor. It is
determined by how well the entangled state is formed and detected.
In our case it can be close to one since it has been shown that such
state can be prepared with a fidelity of nearly $95~\%$. $N$ in this
case is $2$ since two ion maximally correlated state is used for the
measurement. Considering same $f$ as in single ion
experiment~\cite{Nov93}, the figure-of-merit will be $2$ times
higher in present experiment. In other words, it would be possible
to achieve the same precision by performing $1/4^{th}$ no. of
experiments as compared to that on single ion. It can in principle
be further improved by considering the correlated state of more than
two ions.

The size of the PNC light shift could in principle be increased by
increasing the amplitude $E'_{0}$. However, as the amplitude is
increased, the off-resonant couplings become more and more important
which effectively deteriorate the coherence of the entangled state.
The induced loss rate is~\cite{Tim03}
\begin{equation}
\label{eqn9} \Gamma^{loss}_{\gamma j m} =
\frac{e^{2}}{4\hbar^{2}}\sum_{\gamma',m',\pm\omega}\frac{|\langle\gamma'j'm'|\textbf{E}.\textbf{r}|\gamma
j
m\rangle|^{2}}{(\omega_{\gamma'}-\omega_{\gamma}\pm\omega)^{2}}\frac{\omega^{3}}{(\omega_{\gamma'}-\omega_{\gamma})^{3}}\Gamma_{\gamma'j'},
\end{equation}
where $\Gamma_{\gamma'j'}$ is the spontaneous transition rate out of
$|\gamma'j'\rangle$. Considering the off-resonant coupling from 6P
levels $\Gamma^{loss}_{\gamma j m}$ for the states
$|6S_{1/2},m=1/2\rangle$, $|5D_{3/2},m=1/2\rangle$ and
$|5D_{3/2},m=3/2\rangle$ of Ba$^{+}$ has been estimated for an
electric field $E'_{0}=1.6\times10^{6}$ V/m and it turns out to be
$0.0044$, $0.007$ and $0.0008$ Hz respectively. The total induced
loss rate is comparable to the natural decay rate of
5D$_{3/2}$~($0.012$~Hz) and hence the electric field amplitude
mentioned earlier is the maximum for this experiment. The
off-resonant light shift for such a larger electric field is also
significant, but for a linear polarization the ground state Zeeman
sublevels suffer scalar shift and it does not change the Larmor
frequency. However, in presence of small circular polarization in
$E'$ laser the sublevels experience a vector shift~\cite{Sta06} that
can mimic PNC measurement. This systematic can be measured by
performing the same experiment described above but in absence of
$E''$ laser so that there is no interference. Alternatively, the
ions could be placed at the antinodes of $E'$ laser while one of
them at the node of $E''$ laser. A small magnetic field gradient
along the trap axis is a major source of systematic but it can also
be eliminated by repeating the experiment by exchanging the role of the two ions or on the
same ion in absence of the laser fields. In order to finally extract
the PNC induced $E1$ amplitude it is necessary to know the electric
field at the ion position $E'_0(0)$, $E''_0(0)$ and the quadrupole
light shift. The electric fields could be measured by off-resonant
excitations since the related matrix elements for Ba$^+$ are well
known~\cite{Sah09}. The quadrupole light shift can as well be
measured by using the technique of generalized Ramsey interference
experiment~\cite{Ram85,Hae03}. Using two ions instead of one ion in
a linear ion trap may lead to unwanted stray electric field which is
a major concern for parity mixing. Since the ions are side band
cooled to the ground state of their COM mode, the field at the ion
equilibrium position must be zero. The ions in a linear string of
Coulomb crystal have a wavepacket span which is negligible as
compared to the wavelength of the standing wave. Therefore they can
be considered to be at rest. The first order effects due to stray
fields as well as the trapping potential are not only displaced from
the PNC transition by multiples of trap frequency but are also
negligibly small due to sideband cooling.

In case of non-zero nuclear spin isotopes the Physics of PNC is even
richer because of the presence of a tiny nuclear spin dependent
contribution. The measurement of the nuclear spin dependent (NSD)
part and hence the nuclear anapole moment in $E1_{PNC}$ appears to be difficult by driving RF spin flip
transition on a single ion but it is feasible with the technique
described here. For example, in spin $ I=3/2$ isotopes there is only
one $M1$ allowed transition (between $m_{F}=1, 0$ of $F=2, S_{1/2}$
in presence of laser fields connecting $F=2, S_{1/2}$ and $F'=3,
D_{3/2}$) where the quadrupole transition induced light shift does
not change the Larmor frequency. This is essential for measuring the
total PNC light shift. In order to extract the NSD part in
$E1_{PNC}$, other transitions of the same isotope need to be
considered to measure the light shift due to the total PNC
amplitude. The Larmor frequency between those two levels contains
not only the differential PNC shift but also the differential
quadrupole light shift which is a serious systematic effect. However, two
entangled states (Eq.~\ref{eqn7}) with
$|1\rangle_{i}=|m_{F}=2,F=2,6S_{1/2}\rangle_{i}$ ,
$|0\rangle_{i}=|m_{F}=-2,F=2,6S_{1/2}\rangle_{i}$ and
$|1\rangle_{i}=|m_{F}=1,F=2,6S_{1/2}\rangle_{i}$,
$|0\rangle_{i}=|m_{F}=-1,F=2,6S_{1/2}\rangle_{i}$ can be formed to
measure the NSD contribution. The measured light shifts with these
two states contain both NSD and NSI parts which are same in two
transitions but multiplied by associated Clebsch-Gordan
coefficients. It is therefore, convenient to separate out both
contributions with high precision.

We have shown that a two ion entangled state is a better tool for
the measurement of parity violating light shift as compared to the
single ion experiment. Various systematics present in a single ion experiment are absent in this case and some of them can even be measured in this case. The statistical
signal-to-noise ratio can be improved with this maximally correlated
state. The measurement of nuclear spin dependent contribution and
nuclear anapole moment is feasible using correlated atomic states as shown here. The
experimental techniques involved here are regularly in use by the
quantum computation community. Therefore it is feasible with today's
technology.

\noindent\textbf{Acknowledgements.}

The authors would like to thank Prof. B. P. Das at IIA, Bangalore
and Dr. C. F. Roos at IQOQI, Innsbruck for fruitful discussions.
P.~Mandal acknowledges the financial support from CSIR, India. We
acknowledge the financial support of SERC, DST.

\end{document}